\documentclass[12pt]{iopart}

\usepackage{booktabs}
\usepackage{color}
\usepackage{mathrsfs}
\usepackage{ulem}
\usepackage{graphicx}
\usepackage{multirow}
\usepackage{makecell}
\usepackage{longtable}
\usepackage{diagbox}
\usepackage{slashbox}
\usepackage{url}
\usepackage{hyperref}

\renewcommand\thesection{\arabic{section}}
\newcommand{\minitab}[2][l]{\begin{tabular}{#1}#2\end{tabular}}

\def\beq{\begin{equation}}
\def\eeq{\end{equation}}

\begin{document}

\title[]{How to locate the QCD phase boundary by scanning observable in the phase plane }

\author{Yanhua Zhang$^1$,Xue Pan$^2$,Lizhu Chen$^3$,Mingmei Xu$^1$,Zhiming Li$^1$,Yeyin Zhao$^1$,Yuanfang Wu$^1$}

\address{$^1$ Key Laboratory of Quark and Lepton Physics (MOE) and
Institute of Particle Physics, Central China Normal University, Wuhan 430079, China}

\address{$^2$ School of Electronic Engineering, Chengdu Technological University, Chengdu 611730, China}

\address{$^3$ School of Physics and Optoelectronic Engineering, Nanjing University of Information Science and Technology, Nanjing 210044, China}
\ead{\mailto{zhangyanhua@mails.ccnu.edu.cn}, \mailto{wuyf@mail.ccnu.edu.cn}}

\vspace{10pt}

\begin{abstract}\\
For small volume of the quark-gluon plasma formed in heavy ion collisions,
the observable near criticality must obey finite-size scaling. According to the finite-size scaling, there exists a fixed point at the critical temperature, where scaled susceptibility at different system sizes intersect. It also exists at the transitional temperature of a first order phase transition and can be generalized to the region of the crossover. In order to quantify the feature of the fixed point, we introduce {\it the width of a set of points}. When all points in the set are in their mean position within error, the width reaches its minimum, and all points merge into the fixed point. Using the observable produced by the Potts model, we demonstrate that the contour plot of the width defined in this study clearly indicates the temperatures and exponent ratios of fixed point, which could correspond to either the critical point, the points on the transition line, or the crossover region. This method is therefore instructive to the determination of QCD phase boundary by beam energy scan in relativistic heavy ion collisions.

\end{abstract}

\hspace{1.7cm}Keywords: QCD phase transition, finite Size Scaling, fixed Point,three-\\
\hspace{2cm}dimensional three-state Potts model

\maketitle
\section{Introduction}

The phase transition from hadron gas to quark-gluon plasma (QGP) is one of the most fundamental
properties of Quantum Chromodynamics (QCD). The QCD phase diagram is expected to be a first-order phase transition at low temperature $T$ and large baryon chemical potential $\mu_B$~\cite{first-order-a,first-order-b,first-order-c,first-order-d,first-order-e,first-order-f}. The transition line ends at a critical end point (CEP), which belongs to the 3D Ising universality class~\cite{Ising-a,Ising-b,scan-a}. Nearby the region of $\mu_B=0$, the calculation of lattice QCD has shown it is a crossover~\cite{nature-crossover}. However, the position of the phase boundary has not been determined by either calculations of lattice QCD or experiment.

To map the QCD phase diagram from experimental side, beam energy scan
(BES) are suggested and in progress at relativistic heavy-ion collisions (RHIC), FAIR, and NICA~\cite{scan-a,scan-b,scan-c,scan-d}. Varying the incident energy of the collisions ($\sqrt{s_{NN}}$),
the $T$ and $\mu_B$ of formed system change in the phase plane accordingly.  At a given $\sqrt{ s_{NN}}$, the associated $T$ and  $\mu_B$ of freeze-out curve are usually estimated by the
phenomenology of thermal models~\cite{obversables-scaling-a,obversables-scaling-b,center-energy}. So the beam energy scan in fact tune the $T$ and/or $\mu_B$ of formed system in the phase plane~\cite{obversables-scaling-a,obversables-scaling-b,scan-b,S-T-u}.

As we know, in the thermodynamic limit, i.e., an infinite number of particles and volume, the
phase transition is characterized by singularity in the derivatives of the
thermodynamic potential, e.g., the specific heat and susceptibility ($\chi_2$).
Discontinuities in the first and second derivatives signal the first and second order phase
transitions, respectively. The generalized susceptibilities ($\chi_i$) are
theoretically calculable by Lattice QCD~\cite{cumulant-lattice-a,cumulant-lattice-b}. Their
corresponding cumulants of the conserved charges and correlations are experimentally
measurable, and considered as the sensitive observable of the critical fluctuations~\cite{cumulant-a,cumulant-b,correlation-koch,correlation-Lin} .

The non-monotonic dependence of those cumulant on incident energy is
considered as an indicator of the CEP of QCD phase
transition\cite{star-prl-1,star-prl-2}. But they are not specific to the critical
point, and may also appear at other points of the first-order phase transition line and the crossover region~\cite{Xue-JPG42, Zhonghs-Luoxf-2016}. So non-monotonic fluctuations are not sufficient in concluding a CEP.

Due to the small volume of the QGP formed in heavy ion collisions, a possible CEP will be blurred.
Critical related fluctuations will be severely influenced by the finite volume.
The singularities of $\chi_i$ are smeared into finite peaks with modified positions and widths~\cite{finite-size-a,finite-size-b}. With decrease of the volume size, the position of CEP shifts towards smaller temperatures and larger values of the chemical potential~\cite{brazil, PRD90-2014, PLB713-2012}. The peak position indicates the so called pseudo-critical point. For a restricted volume which is not very small, the CEP has to be determined by the finite-size scaling of the observable~\cite{Wu-Lizhu,Lizhu-wu,Xue-JPG42, obversables-scaling-a,obversables-scaling-b}.

In statistical physics, the critical related observable is usually the function of temperature and system size, such as the order parameter $m(T,L)$. In the vicinity of the critical temperature $T_{\rm C}$, the order parameter follows the finite size scaling (FSS)~\cite{shift-b,scaling},
\vspace{1.3mm}
\begin{equation}\label{scaling function 1}
m(T,L)=L^{-\beta/\nu}f_{m}(t L^{1/\nu}),
\end{equation}
\vspace{1.6mm}
cf., Fig.~1(a). Where $t=(T-T_C)/T_C $ is reduced temperature. $L$ is the system size. $f_{m}$ is scaling function. $t L^{1/\nu}$ is scaled variable. $\beta$ is the scaling exponent of order parameter. $\nu$ is the scaling exponent defined by the divergence of the correlation length $\xi\propto |\tau|^{-\nu}$. The scaling exponents characterize the universal class of the phase transition. The exponent ratio $\beta/\nu$ is usually a fraction between the spatial dimension $d$ and zero.

In relativistic heavy ion collisions, the scaling behaviour of several critical related observable has been studied~\cite{obversables-scaling-a,obversables-scaling-b, RoyPRL, Roy1, Roy2,PRD97-NA49}. For example, the finite size scaling of intermittency analysis for transverse momentum in central A+A collisions from NA49 experiments recently restrict the CEP to the region 119 MeV $ \le T^{\rm CEP}\le $ 162 MeV, and 252 MeV $\le \mu_B^{\rm CEP}\le $ 258 MeV~\cite{PRD97-NA49}.The scaling of emission source radii difference in Au+Au collisions at RHIC BES energies was roughly showed at $T^{\rm CEP}=165$ MeV and $\mu^{\rm CEP}_{B}=95$ MeV~\cite{RoyPRL}.

On those analyses, the critical temperature ($T_C$) and exponent of Ising universality are usually assumed in advance. Then, the observable around given $T_C$ are plotted in the plane of scaled variable and observable. Scaling behaviour are usually judged by the naked eye at the end. As the scaling function is unknown in priori, the whole processes cannot be quantified by a usual best $\chi^2$ fit. Its uncertainty and precision are therefore difficult to estimate.

Moreover, how the observable approach its critical value with change of temperature and exponent is not demonstrated in the plot. Whether the chosen $T_C$ and exponent are the best parameters for the scaling are not clear. If it is the first order phase transition line and crossover region has not been discussed exclusively either.

In fact, the feature of fixed point associated with Eq.~(\ref{scaling function 1}) of FSS is helpful in those aspects. At critical temperature $T=T_{C}$, the scaled variable keeps zero, independent of system size, all curves of scaled observable ($m (T,L)L^{\beta/\nu}$) at different system sizes versus temperature ($T$), instead of scaled variable ($t L^{1/\nu}$), intersect to {\rm a fixed point} of size scale invariance, cf., Fig.~1(b), i.e., the fixed point under the transformation of renormalized group~\cite{scaling, Lizhu-Wucpod11}.

From Fig.~1(b), the critical temperature is indicated by the position of the fixed point. How the observable at different system sizes approach the critical one with change of temperature are clearly demonstrated, contrast to Fig.~1(a). So if the behaviour of fixed point in the plot can be well quantified, the location of the critical point will be precisely given out.

\begin{figure}
\centering
\includegraphics[width=0.67\linewidth]{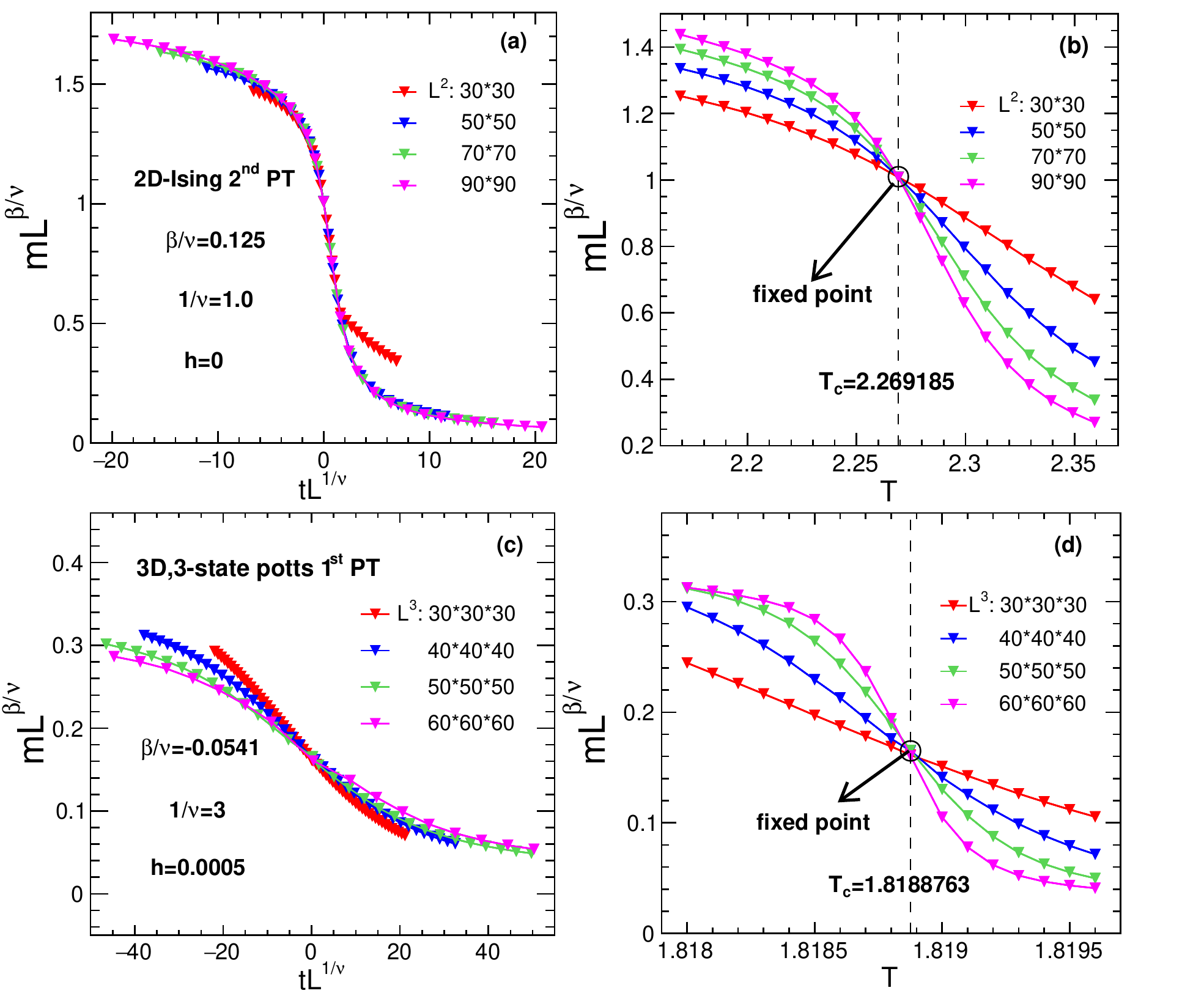}
\caption{\label{Fig. 1} The mean of order parameter times the powers of system size ($m L^{\beta/\nu}$) versus scaled variable ($tL^{1/\nu}$) (left column (a) and (c)), and the temperature $T$ (right column (b) and (d)) for the second (top panels) and the first (bottom panels) order phase transitions from 2D Ising and 3D three-state Potts models at external fields $h=0$ and $h=0.0005$, respectively.}
\end{figure}

It is known that the finite size scaling keeps valid on the first order phase transition line~\cite{scaling}, cf., the bottom panels of Fig.~1, which is also obtained by the transformation of renormalization group ~\cite{Fisher:1982xt,vanLeeuwen:1975zz,PhysRevLett.35.477}. Here the exponent ratio is usually an integer.

In contrast to the first and second order phase transitions, the fixed point disappears in the crossover region. The observable is system size independent~\cite{nature-crossover}. All kinds of observable changes smoothly without singularity. If we still scale the observable in general scaling form, the exponent ratio is zero.

The fixed points corresponding to the second and first order phase transitions are showed in Fig.~1 (b) and (d). They come from the 2D Ising and 3D three-states Potts models, respectively~\cite{2D-ising-a,2D-ising-b,2D-ising-c,potts-model-wu,Xue-JPG42}. Both of them show the feature of fixed point, i.e.,
all size curves intersect at critical (transition) temperature. Any deviation from the critical (transition) temperature, the points with different system sizes would go away from each other. So we can define
the width of the set of points to quantify the feature of fixed point.

In this paper, how to locate the fixed point by the contour plot of defined width is demonstrated. In Section II, the feature of fixed point is defined by the width of a set of points. Then in Section III, using the 3D three-state Potts model, the samples at three external fields, which correspond to the CEP, the first order phase transition, and crossover region, are produced respectively. In section IV, how to find the position of the fixed point in a given sample is demonstrated. A brief summary and conclusions are presented in Section V.

\section{The description of fixed point}

In general, the finite-size scaling Eq.~(\ref{scaling function 1}) is valid for the observable which is a phase transition related fluctuations. If we denote the observable as $Q(T,L)$, its finite-size scaling would be,
\begin{equation}\label{observable scaling}
   Q(T,L)=L^{\lambda/\nu}f_Q(tL^{1/\nu}).
\end{equation}
\noindent Here $\lambda$ is the scaling exponent of the observable $Q(T,L)$. $f_Q$ is scaling function with the scaled variable $tL^{1/\nu}$. Multiplying $L^{-\lambda/\nu}$, Eq.~(\ref{observable scaling}) becomes,
\begin{equation}\label{observable scaling function}
f_Q(t L^{1/\nu})=Q(T,L)L^{-\lambda/\nu}.
\end{equation}

\noindent It implies a scaling plot, scaled observable $Q(T,L)L^{-\lambda/\nu}$ versus scaled variable $t L^{1/\nu}$, where all curves at different system sizes overlap into a single curve nearby the critical temperature, as illustrated in Fig.~1(a).

At critical temperature $T=T_{\rm C}$, the scaled variable $tL^{1/\nu}=0$, independent of system size $L$, and the scaling function becomes a constant, i.e.,
\begin{equation}\label{scaling function zero}
f_Q(0)=Q(T_{\rm C},L)L^{-\lambda/\nu}.
\end{equation}

This means the existence of {\it a fixed point} in the plot of $Q(T,L)L^{-\lambda/\nu}$ versus the temperature $T$, instead of scaled variable $t L^{1/\nu}$. The curves of different system sizes are intersected to a fixed point at $T_{\rm C}$, and separated from each other when $T$ deviates from $T_{\rm C}$, as shown in Fig.~1(b).

In Fig.~1(b), at a given temperature, the collection of the points of different system sizes can be defined as a set. When $T$ approaching $T_{\rm C}$, all points in the set becomes closer and closer to each other. The trend of all points going close to each other is necessary for forming a unique intersect point. In order to quantify the relative distance between the points in the set, we define the width of all points in the set as the variance of all size points ($\Delta S$) to their mean position, i.e.,
\begin{equation}\label{define D}
 D(T,a)=\sqrt{\frac{\Delta S_{Q(T,L)L^{a}}}{N_L-1}}.
\end{equation}
Here, $N_L$ is the number of different sizes. $\Delta S_{Q(T,L)L^{a}}$ is the error weighted variance of all size points to their mean position, i.e.,
\begin{equation}\label{define D 1}
 \hspace{-5mm}     \Delta S_{Q(T,L)L^{a}}=\sum_{i=1}^{N_L}\frac{[Q(T,L_i)L^{a}_i-\langle Q(T,L)L^{a}\rangle]^2}{\omega^2_i}.
\end{equation}
$\omega_i = \delta [Q(T,L_i)L^{a}_i]$ is the error of $Q(T,L_i)L^{a}_i$. $\langle Q(T,L)L^{a}\rangle$ is the weighted mean, i.e.,
\begin{equation}\label{define D 2}
\langle Q(T,L)L^{a}\rangle = \frac{\sum_{i=1}^{N_L}Q(T,L_i)L^{a}_i /\omega^2_i}{\sum_{i=1}^{N_L}1/\omega^2_i}.
\end{equation}

Since all points in the set are at a given temperature, and the variance $\Delta S_{Q(T,L)L^{a}}$ is all points with different system sizes to their mean, the width $D(T,a)$ is therefore only the function of temperature $T$ and exponent ratio $a=-\lambda/\nu$, which is usually unknown and varies with observable.

Such defined width $D(T,a)$ describes the relative distance of all points to their mean position. At the critical temperature and exponent ratio, all points within error are coincident to their mean position, which is closest to the theoretically expected fixed point of size scale invariance. $D(T,a)$ reaches its minimum, around unity, which is similar to that of $\chi^2$ in curve-fitting. Where the variance is the measured points to a given curve. The minimum of $\chi^2$ defines the best fitting of all points to the curve. When the temperature deviates from the critical one, the points of different system sizes go away from each other, and the width becomes larger.

In real experiments, due to the error of the observable and related uncertainties, the fixed point may not converge to an ideal point, and $D(T,a)$ may be larger than unity. Nevertheless, if there is a fixed point in $T$ and $a$ plane, the $D(T,a)$ will change with $T$ and $a$ and converge to a minimum. This is essential for forming a fixed point.

\begin{figure}[!htb]
\centering
\includegraphics[width=0.68\linewidth]{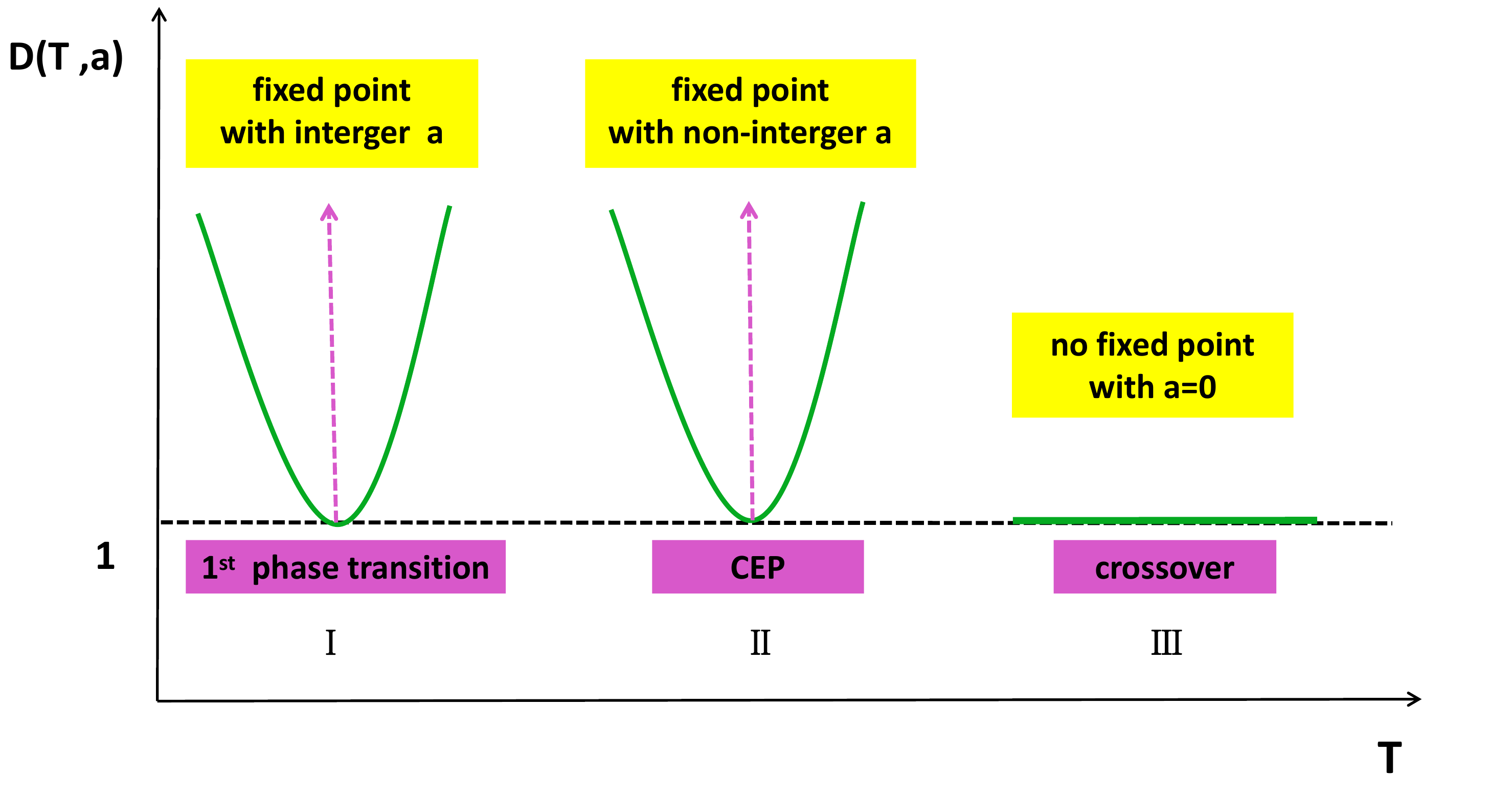}
\caption{\label{Fig. 2} (Color line) $D(T,a)$ nearby the temperatures of the first (I), and the second (II) order phase transitions, and crossover region (III).}
\end{figure}

At three different regions of phase boundary, the change of $D(T,a)$ with $T$ and $a$ are expected to be the cases of I, II and III, respectively, as showed in Fig.~2.

In the case of I, the temperature is low. $D(T,a)$ has a minimum at phase transition temperature, where the ratio $a$ is around $(n-1)d$, an integer. $n$ is the order of susceptibility. It characterizes the fixed point of the first order phase transition.

In the case of II,  the temperature is in the middle. $D(T,a)$ has also a minimum at the critical temperature, and ratio $a$ is a fraction, in contrast to the case of I. It indicates the fixed point of the second order phase transition, i.e., the CEP.

In the case of III, the temperature is high. $D(T,a)$ keeps in the same minimum at all temperature, and ratio $a$ is around zero. This implies $D(T,a)$ is a constant, independent of the temperature. The observable is the system size independent. It is the feature of crossover region.

In the following, it is interesting to display how the minimum of defined width can locate the fixed point from a sample of observable which contains the CEP, the point on the first transition line, and crossover region, respectively.

\section{Three specific samples generated by the Potts model}

The 3D three-state Potts model is one of the standard paradigms of lattice
QCD~\cite{potts-Z-a,potts-Z-b,potts-Z-c,potts-Z-d}. It is a pure gauge QCD effective model,
and shares the same Z(3) global symmetry as that of QCD with finite temperature and infinite
quark mass. Where external magnetic field plays the role of the quark mass of the finite
temperature QCD. At vanishing external field, the temperature-driven phase transition has
proved to be of the first-order~\cite{potts-first-order-a, potts-first-order-b}. With increase
of the external field, the first-order phase transition weakens and ends at the critical point
$(1/T_{\rm C}, h_{\rm C}) = (0.54938(2), 0.000775(10))$, which also belongs to 3D Ising
universality class~\cite{potts-Z-b,potts-ising-b}. Above the critical point, it is a
crossover region.

The phase structure of the Potts model is comparable with that of deconfinement and chiral
symmetry restoration in QCD, which both have the line of first-order
phase transition, and end in a second-order endpoint, cf., Fig~1(b) in ref~\cite{Xue-JPG42}.
The baryon chemical potential in QCD acts as the external field in the Potts model.
So applying the defined width to the sample generated by the Potts model is instructive to
that in relativistic heavy ion collisions.

The 3D three-state Potts model is described in terms of spin variable
$s_i \in {1, 2, 3}$, which is located at sites $i$ of a cubic lattice of size $V = L^3$.
The Hamiltonian of the model is defined by~\cite{Xue-JPG42,potts-ising-b},
\begin{equation}\label{H}
H=\beta E-hM.
\end{equation}
\noindent The partition function is,
\begin{equation}\label{Z}
Z(\beta, h) =\sum_{\{{s}_{i}\}}e^{-(\beta E-hM)}.
\end{equation}
Where $\beta=1/T$ is the reciprocal of temperature, and $h=\beta H$ is the normalized external magnetic field. $E$ and $M$ denote the energy and magnetization respectively, i.e.,
\begin{equation}\label{E and M}
E=-J\sum_{\langle i,j\rangle}\delta(s_i,s_j), \ {\rm and} \ M=\sum_{i}\delta(s_i,s_g).
\end{equation}
Here $J$ is an interaction energy between nearest-neighbour spins $\langle i,j\rangle$,
and set to unity in our calculations. $s_g$ is the direction of ghost spin for the magnetization
of non-vanishing external field $h>0$. For vanishing external field the model is
known to have a first order phase transition. With increase of the external field, the first-order
phase transition line ends at a critical point $(\beta_{\rm C},h_{\rm C})$.

The order parameter of the system is defined as,
\begin{equation}\label{m}
m(T,L)=\frac{3\langle M\rangle}{2V}-\frac{1}{2}.
\end{equation}
However, at critical point $(\beta_{\rm C},h_{\rm C})$ the original operators $E$ and $M$ lose their meaning as temperature-like and H-like, i.e., symmetry breaking couplings, as those in Ising model. The order parameter and energy-like observable has to be redefined as the combination of the original $E$ and $M$, i.e.~\cite{potts-ising-b},
\begin{equation}\label{M}
\tilde{M}=M+sE,\ {\rm and } \ \tilde{E}=E+rM.
\end{equation}
The Hamiltonian in terms of $\tilde {M}$ and $\tilde {E}$ is,
\begin{equation}\label{H}
H=\tau \tilde{E}-\xi \tilde{M}.
\end{equation}
Where the new couplings are given by,
\begin{equation}\label{eq-refer}
\xi = \frac{1}{1-rs}(h-r \beta),\ {\rm and }\ \tau = \frac{1}{1-rs}(\beta-sh).
\end{equation}
Here $r$ and $s$ are the mixing parameters and have been determined in ref. ~\cite{potts-ising-b} by,
\begin{equation}\label{r and s}
r^{-1}=(\frac{d \beta_{\rm C}(h)}{d h})_{h=h_{\rm C}},\ {\rm and} \ \langle\delta \tilde{M}\cdot \delta \tilde{E}\rangle =0,
\end{equation}
with $\delta \tilde{X}=\tilde{X}-\langle\tilde{X}\rangle$ for $X=M$, or $E$.

The order parameter in terms of $\tau$ and $\xi$ is,
\begin{equation}\label{m and M}
m(\tau,\xi)=\frac{1}{L^3}[\tilde M(\tau,\xi)-\langle \tilde M(\tau_{\rm C},\xi_{\rm C})\rangle].
\end{equation}
It is the most sensitive observable to the phase transition.
According to Eq.~(\ref{r and s}) and ~(\ref{m and M}), it can be written in terms of $T$ and $h$ as,
\begin{equation}
 m(T,h)= \frac{1}{L^3}[\tilde M(T,h)-\langle \tilde M(T_c,h_c)\rangle].
\end{equation}
Where $\tilde M(T,h)=M_L(T,h)+sE(T,h)$, and $M(T,h)$ is obtained by Eq.~(\ref{E and M}) from generated spins at lattice. The mixing parameter $s$ is estimated from the table 2 of ref.~\cite{potts-ising-b} and the Eq.~(\ref{r and s}).

In the following, the observable will be taken as the mean of absolute order parameter, i.e.,

We performed simulations at three fixed external fields, $h=0.0005$, $0.000775$, 0.002 and $18$
$T$-values starting from $T_0=1.8180$ with $\Delta T=0.0001$, which covers the points on
the first order phase transition line, the CEP, and crossover region, respectively. The temperature and exponent ratio at three special points are listed in the bracket of Table I. The sample is generated for four system sizes $L=30,\ 40,\ 50,\ 60$. The $m(T,L)$ at different temperatures and system sizes are obtained in total 100,000 configurations.

\section{Locating the fixed point by defined width}

Now, suppose that we have three samples. In each of them, only the mean of
the order parameter for different temperatures and system sizes are presented. All other information is unknown. Let's apply only these known information to calculate the width $D(T, a)$ and see if we can find the position of temperature and exponent ratio corresponding to the first and second order phase transitions and crossover region.
\begin{equation}
\hspace{-3.1mm}\langle m(T,h)\rangle=
\langle|\frac{1}{L^3}[\tilde M(T,h)-\langle\tilde M(T_c,h_c)\rangle]|\rangle.
\end{equation}
For fixed external field $h$, it is the function of temperature and system size, i.e.,
$\langle m(T,L)\rangle$.

\begin{table}
\centering
\setlength{\arrayrulewidth}{0.3pt}
\begin{tabular}{|c|c|c|c|}
\hline
\multirow{2}{*}{\diagbox{Sample}{Variable}}&  \multirow{2}{*}{$D_{min}(T,a)$}       &\multirow{2}{*}{$T$}      &\multirow{2}{*}{$a$} \\
                               &                                   &                                          &                               \\
 \hline
 \multirow{3}{*}{$2^{\rm nd}$ order PT}&\multirow{3}{*}{1.0291$\pm$0.2946} &\multirow{3}{*}{\minitab[c]{1.82023\\(1.82023372)}}&\multirow{3}{*}{\minitab[c]{0.583\\(0.564)}}\\
                               &                                                                 &                         &                      \\
                               &                                                                 &                         &\\

\hline
 \multirow{2}{*}{\minitab[c]{$1^{\rm st}$ order PT}}&\multirow{2}{*}{1.5287$\pm$0.5591}&\multirow{2}{*}{\minitab[c]{1.81887\\(1.8188763)}}&\multirow{2}{*}{\minitab[c]{-0.047\\(-0.0541)}}\\
                             &                                 &                             &       \\
\hline
 \multirow{3}{*}{\minitab[c]{crossover}}&\multirow{3}{*}{\minitab[c]{$\sim 1$ for all $T$}}&\multirow{3}{*}{1.82585039}&\multirow{3}{*}{-0.1$\sim$0.1}\\
                             &                                 &                             &       \\
                            &                                 &                             &       \\
\hline
\end{tabular}
\caption{$T$ and $a$ at $D_{\rm min}(T, a)$ and $T_{\rm C}$ and $a_{\rm C}$ in the Potts model (inside brackets) for three samples.}
\end{table}

According to defined width $D(T, a)$ in Eq.~(\ref{define D}), the corresponding width of the mean of  order parameter at a given temperature $T$ and exponent ratio $a$ is,
\begin{equation}
D(T,a)=\sqrt{\frac{\Delta S_{\langle m(T,L)\rangle L^{a}}}{N_L-1}}.
\end{equation}
Where,
\begin{equation}
   \begin{array}{cl}
 \displaystyle \hspace{-5.5mm}  \Delta S_{\langle m(T,L)\rangle L^{a}}=\sum_{i=1}^{N_L}\frac{1}{\omega^2_i}\times
       \big[\langle m(T,L_i)\rangle L^{a}_i - \langle\langle m(T,L)\rangle L^{a}\rangle\big]^2 \\
    \end{array}
\end{equation}
$\omega_i$ is the error of  $[\langle m(T,L_i)\rangle L^{a}_i]$, and
\begin{equation}
\langle\langle m(T,L)\rangle L^{a}\rangle = \frac{\sum_{i=1}^{N_L}\langle m(T,L_i)\rangle
L^{a}_i /\omega^2_i}{\sum_{i=1}^{N_L}1/\omega^2_i}
\end{equation}
is error weighted average. Here, the summation number $N_L$ equals to 4 for four system sizes $L=30,\ 40,\ 50,\ 60$.

The exponent ratio $a=-\lambda/\nu$ is unknown parameter, it can be
$-\infty$ to $\infty$ in principle. For a given temperature, we can tune the ratio to see when it
makes $D(T, a)$ minimum. So we present contour of $D(T, a)$ in the plane of $T$ and $a$
for three samples in Fig.~3(a), (b) and (c), respectively. Where colour bar
on the right side indicates the values of $D$. The red and blue zones are minimum and maximum,
respectively. The range of $a$ is from $-1.2$ to $1.15$ with the interval 0.05.

For the sample with the external field, $h=0.000775$, the contour lines in Fig. ~3(a) show that $D(T, a) $ gradually converges to a minimum red area, where the temperature and ratio ranges are respectively $1.8200\sim 1.8204$ and $0.4\sim 0.65$. This implies that the width of all points at the same temperature with different system sizes converges to a minimum at a specified $T$ and $a$. It is the  nature of fixed point.

\begin{figure}[h]
  \centering
  \noindent\makebox[\textwidth][c] {
    \includegraphics[width=0.87\paperwidth]{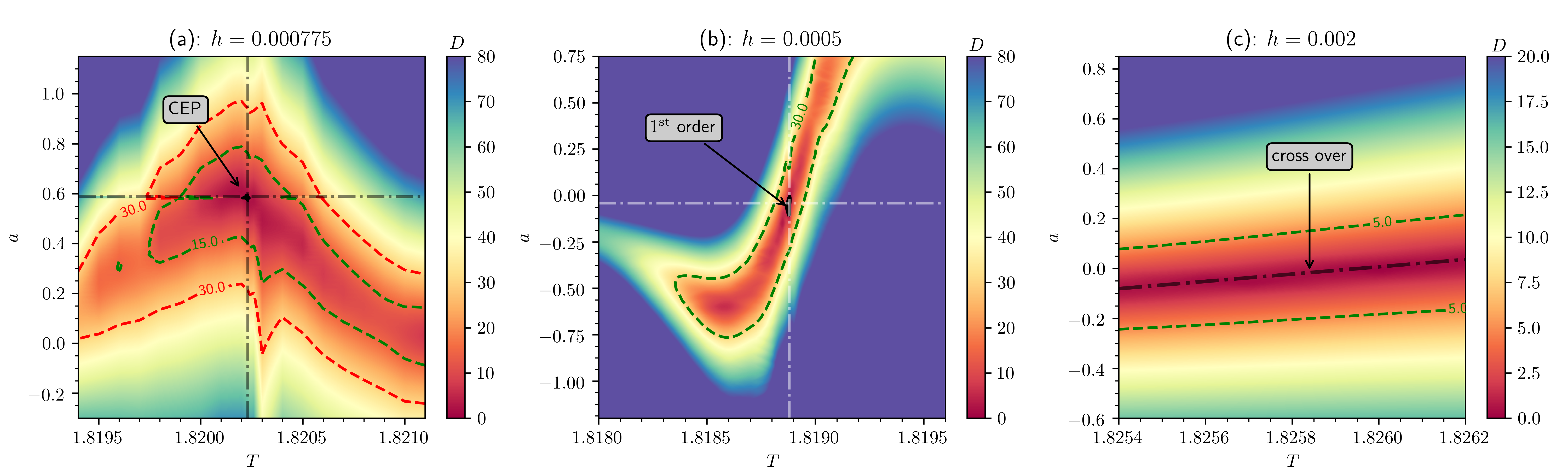} }
\caption{\label{Fig. 3} $D(T, a)$ for three samples with external fields $h=0.000775$ (the CEP)(a),  $0.0005$ (the first order phase transition line) (b),  and 0.002 (crossover region) (c). Dash point lines indicate the coordinates of $D_{\rm min}(T, a)$ and dash lines are isolines.}

\end{figure}

To amplify the fine structure and minimum nearby the red area, we project
$D(T, a)$ to $a$ and $T$ axis, respectively, as showed in Fig.~4(a) and 4(c). In Fig.~4(a),
for a given $T$, there is an $a$ which makes the $D(T, a)$ minimum. Among these lines,
the minimum is the red one with $D_{\rm min}=1.0291\pm  0.2946$, and corresponding $T=1.82023$
and $a=0.583$. They are very close to the original critical temperature (1.82023372) and ratio
(0.564) from the given sample, cf., Table I.


From the projection along the direction of temperature as showed in Fig.~4(c), for a given
ratio $a$, there is also a minimum $D(T, a)$. Among them, the smallest minimum is the red line which corresponds to the same critical temperature and exponent ratio as those from Fig.~4(a).

These features of $D(T, a)$ are consistent with those of the critical point as showed in
the case II of Fig.~2. The contour plot of $D(T, a)$ demonstrate how all curves of the mean of absolute order parameter at different system sizes intersect at the critical temperature and exponent ratio.

Then turn to the sample for the external field $h=0.0005$, its contour plot is showed in
Fig.~3(b). Here, it shows again that $D(T, a)$ gradually converges to a minimum red area. This
means that all curves of scaled observable at different system sizes are more and more close
to each other with the change of $T$ and $a$. The minimum red area corresponds to the ranges of temperature and exponent ratio $1.81885\sim 1.81889$, and $-0.15\sim 0$, respectively.

\begin{figure}
\centering
\includegraphics[width=4.3in]{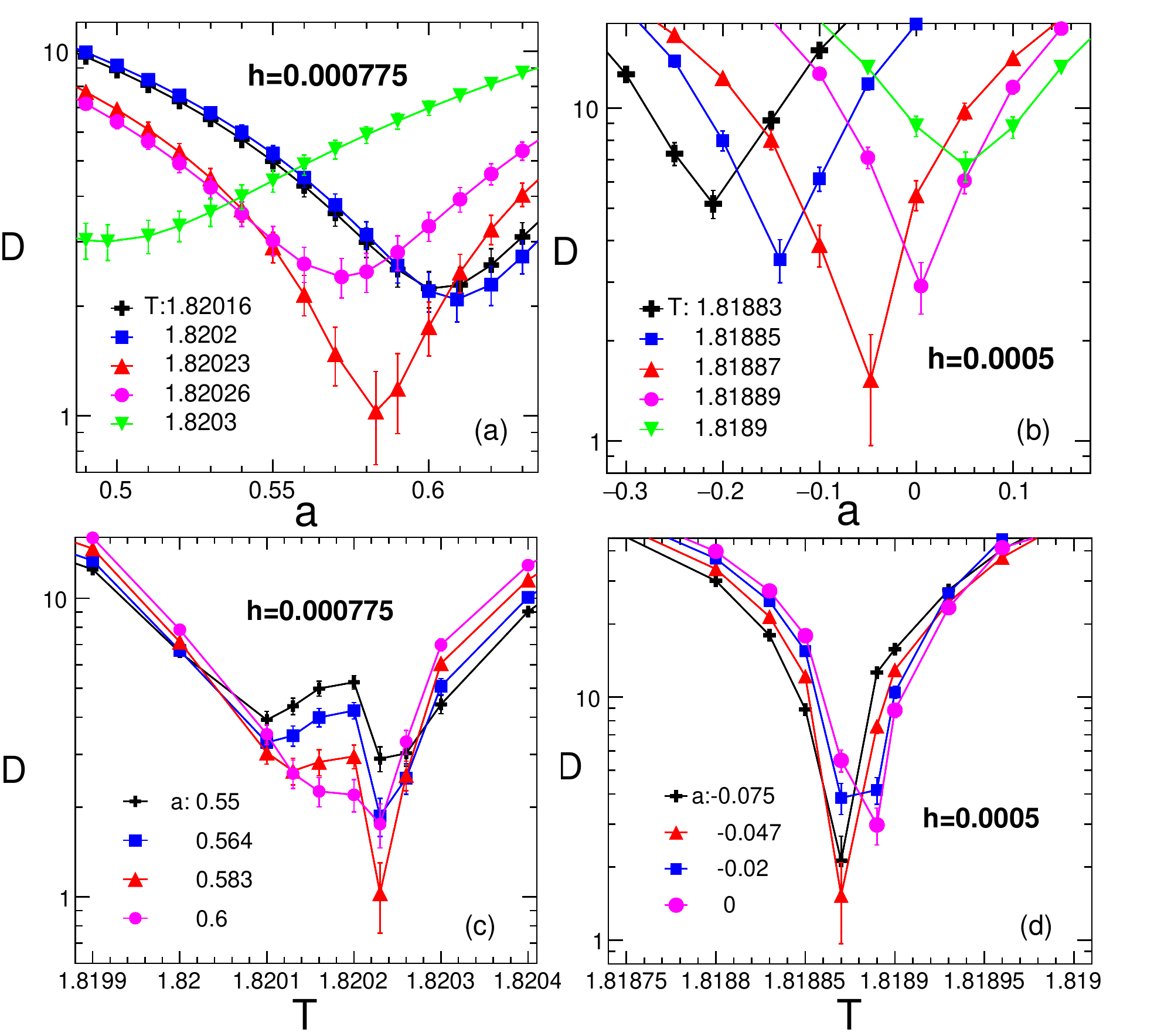}
\caption{\label{Fig. 4} Projections of $D(T, a)$ on  exponent ratio (upper row) and temperature (lower row) axis nearby the critical point (a) and (c), and the point on the first order phase
transition line (b) and (d). }
\end{figure}

The projection plots of $D(T, a)$ versus $a$ for different $T$ are shown in
Fig.~4(b). The curve which has the smallest minimum is the red line with
$ D_{\rm min}(T, a)=1.5287 \pm 0.5591 $, where the temperature is $T=1.81887$, which is
very close to the original one 1.8188763, cf., Table I. The exponent ratio
is -0.054, which is almost the same close to zero as that from the original sample $-0.047$,
cf., Table I. Here the order parameter can be considered as the first order of susceptibility. So the exponent ratio is zero at the first order phase transition line, the same as that for the crossover region.

The projection along the direction of $T$ is showed in Fig.~4(d), for a given
ratio $a$, there is also a minimum $D(T, a)$. The minimum gives the same critical temperature and exponent ratio as those along the direction of $a$ in Fig.~4(b). So $D(T, a)$ again demonstrates the features of fixed point at the first order phase transition line, as those showed in the case I of Fig.~2.

The contour plot of the sample for the external field $h=0.002$ is shown in Fig.~3(c). In
contrast to Fig.~3(a) and 3(b), the $D(T, a)$ in Fig.~3(c) are band lines parallel to the $T$-axis.
This implies that $D(T, a)$ is independent of temperature, and its value is determined by
the ratio $a$ only. The red band is close to zero. This is the same
characteristics as the crossover region showed in the case III of Fig.~2.

So the contour plot of defined width $D(T, a)$ beautifully quantifies the features of the fixed point. When all curves of scaled order parameter at different system sizes intersect to a fixed point, $D(T, a)$ indeed converges to an expected unity.

Although it should be noticed that due to the error of the observable and uncertainties of related parameters in real experimental settings, the minimum of $D(T, a)$ may be larger than the unity and vary with experiments, what's more important for the formation of a fixed point is the trend that the contour plot of $D(T, a)$ converges to a minimum area.

Here, we present three samples of observable which just pass two specified fixed points and the crossover region, respectively. The contour plot of defined width displays the regular regions, as the isolines indicated in Fig~3(a), (b) and (c), respectively. If the sample of observable does not pass the phase boundary, the plot would vary with its covered phase plane. If it is far away from the phase boundary, the observable are independent either of temperature or system size~\cite{Xue-JPG42}. The plot keeps at its minimum. If it approaches to the phase boundary, or the transition temperature, the plot may show some contour regions which are a part of fig.~3(a), or (b), or (c). Therefore, the contour plot of defined width is helpful for exploring the phase boundary.

In relativistic heavy ion collisions, we measure the critical related fluctuations, such as,
the cumulants of conserved charges at different incident energy and collision centrality. They are observable, similar to the mean of the  order parameter in the Potts model. The incident
energy should correspond to the temperature, and/or baryon chemical potential. The relation between
them is currently given by thermal model~\cite{obversables-scaling-a,center-energy}. The size of formed matter is roughly estimated by the radii of Hanbury Brown Twiss (HBT) interferometry~\cite{HBT-1, HBT-2, HBT-3, RoyPRL, Roy1, Roy2}. When all these relations are reliably set up,
the defined width would be directly applicable to the data analysis at RHIC BES.

It should also mention the fact that Eq.~(\ref{observable scaling}) may not exactly hold for some critical related observable, such as energy density, and specific heat~\cite{Book-phase}. For this kind of observable, additional scaling violating terms are not negligible~\cite{Engels:2011km, Blote_1995}. They are usually the function of system size and temperature. So the fixed point will not appear in the plot of the scaled variable versus temperature. Its behaviour may vary with the observable and associated system~\cite{Engels:2011km, Blote_1995}. The suggested contour plot would not converge to a minimum as those showed in Fig.~3, and change with observable and associated system as well.

Therefore, to apply the method to determine the QCD phase structure from RHIC BES, it is necessary to measure observable which likes to the order parameter, or the susceptibility, such as the cumulants of conserved charges. However, it is usually difficult to know if the observable is analogous with susceptibility or specific heat in advance. So it is helpful to measure as much as possible related observable, and exam their corresponding contour plots respectively.

\section{Summary and conclusions}
Since the volume of QGP formed in heavy ion collisions is small, the fluctuations near criticality should follow the finite-size scaling. Based on the finite-size scaling, the CEP corresponds to a fixed point, where all scaled observable at different system sizes intersect. It also exists on the first order phase transition line and can be generalized to the crossover. Their corresponding exponent ratios are respectively fraction, integer and zero. So the phase boundary can be well identified by the corresponding fixed point in the phase plane.

To quantify the feature of the fixed point, at a given temperature, we define the width of a set of points with different system sizes. It is the square root of the variance of scaled observable. When all points in the set are in their mean position within error, the defined width reaches its minimum, and all points are overlapped in an experimental sense, i.e., fixed point. So the minimum of the width corresponds to the position of the fixed point.

Then using the 3D three-state Potts model, we produce the samples at three external fields, which correspond to the CEP, the first order phase transition, and crossover, respectively. The temperature range of each sample covers the whole phase plane. We demonstrate that the minimum of the contour plot of defined width precisely indicates the temperature and scaling exponent ratio of fixed point presented in three samples, respectively.

Therefore, the contour plot of defined width well quantifies how the observable approaches the temperature and exponent ratio of phase transition. It provides an exact and systematic way to locate the critical point, the first order phase boundary, and the crossover region by scanning the related observable in the phase plane. When the relation between incident energy and the temperature of formed matter is well settled down, and system size of formed matter can be reliably estimated, the application of the method would be straight forward.

\section*{Acknowledgement}

We are grateful to Dr. Xiaosong Chen for drawing our attention to the field. The last author would thank Dr. Xingguo Xu and Yu Zhou for critical proof-reading of the paper.

This work is supported in part by the Major State Basic Research Development Program of China under Grant No. 2014CB845402, the Ministry of Science and Technology (MoST) under grant No. 2016YFE0104800, and the NSFC of China under Grants No. 11521064, 11405088, 11647093.

\section*{References}
\bibliographystyle{iopart-num}
\bibliography{ref}

\providecommand{\newblock}{}
\begin{thebibliography}{10}
\expandafter\ifx\csname url\endcsname\relax
  \def\url#1{{\tt #1}}\fi
\expandafter\ifx\csname urlprefix\endcsname\relax\def\urlprefix{URL }\fi
\providecommand{\eprint}[2][]{\url{#2}}

\bibitem{first-order-a}
Yaffe L~G and Svetitsky B 1982 {\em Phys. Rev. D\/} {\bf 26} 963

\bibitem{first-order-b}
Roberge A and Weiss N 1986 {\em Nucl. Phys. B\/} {\bf 275} 734--745

\bibitem{first-order-c}
Fukugita M, Okawa M and Ukawa A 1989 {\em Phys. Rev. Lett.\/} {\bf 63} 1768

\bibitem{first-order-d}
de~Forcrand P and Philipsen O 2002 {\em Nucl. Phys. B\/} {\bf 642} 290--306
  (\textit{Preprint} \eprint{hep-lat/0205016})

\bibitem{first-order-e}
Ejiri S 2008 {\em Phys. Rev. D\/} {\bf 78} 074507 (\textit{Preprint}
  \eprint{0804.3227})

\bibitem{first-order-f}
Bowman E~S and Kapusta J~I 2009 {\em Phys. Rev. C\/} {\bf 79} 015202
  (\textit{Preprint} \eprint{0810.0042})

\bibitem{Ising-a}
Halasz A~M, Jackson A~D, Shrock R~E, Stephanov M~A and Verbaarschot J~J~M 1998
  {\em Phys. Rev. D\/} {\bf 58} 096007 (\textit{Preprint}
  \eprint{hep-ph/9804290})

\bibitem{Ising-b}
Berges J and Rajagopal K 1999 {\em Nucl. Phys. B\/} {\bf 538} 215--232
  (\textit{Preprint} \eprint{hep-ph/9804233})

\bibitem{scan-a}
Stephanov M~A, Rajagopal K and Shuryak E~V 1998 {\em Phys. Rev. Lett.\/} {\bf
  81} 4816--4819 (\textit{Preprint} \eprint{hep-ph/9806219})

\bibitem{nature-crossover}
Aoki Y, Endrodi G, Fodor Z, Katz S~D and Szabo K~K 2006 {\em Nature\/} {\bf
  443} 675--678 (\textit{Preprint} \eprint{hep-lat/0611014})

\bibitem{scan-b}
Andronic A, Braun-Munzinger P and Stachel J 2009 {\em Phys. Lett. B\/} {\bf
  673} 142--145 [Erratum: Phys. Lett.B678,516(2009)] (\textit{Preprint}
  \eprint{0812.1186})

\bibitem{scan-c}
Becattini F, Bleicher M, Kollegger T, Mitrovski M, Schuster T and Stock R 2012
  {\em Phys. Rev. C\/} {\bf 85} 044921 (\textit{Preprint} \eprint{1201.6349})

\bibitem{scan-d}
Stephanov M~A 2004 {\em Prog. Theor. Phys. Suppl.\/} {\bf 153} 139--156 [Int.
  J. Mod. Phys.A20,4387(2005)] (\textit{Preprint} \eprint{hep-ph/0402115})

\bibitem{obversables-scaling-a}
Fraga E~S, Kodama T, Palhares L~F and Sorensen P 2010 {\em PoS\/} {\bf
  FACESQCD} 017 (\textit{Preprint} \eprint{1106.3887})

\bibitem{obversables-scaling-b}
Fraga E~S, Palhares L~F and Sorensen P 2011 {\em Phys. Rev. C\/} {\bf 84}
  011903 (\textit{Preprint} \eprint{1104.3755})

\bibitem{center-energy}
Cleymans J, Oeschler H, Redlich K and Wheaton S 2006 {\em Phys. Rev. C\/} {\bf
  73} 034905 (\textit{Preprint} \eprint{hep-ph/0511094})

\bibitem{S-T-u}
Tawfik A~N 2014 {\em Int. J. Mod. Phys. A\/} {\bf 29} 1430021
  (\textit{Preprint} \eprint{1410.0372})

\bibitem{cumulant-lattice-a}
Cheng M {\em et~al.\/} 2009 {\em Phys. Rev. D\/} {\bf 79} 074505
  (\textit{Preprint} \eprint{0811.1006})

\bibitem{cumulant-lattice-b}
Stephanov M~A 2009 {\em Phys. Rev. Lett.\/} {\bf 102} 032301 (\textit{Preprint}
  \eprint{0809.3450})

\bibitem{cumulant-a}
Koch V 2010 {Hadronic Fluctuations and Correlations} {\em Chapter of the book ,
  R. Stock (Ed.), Springer, Heidelberg, 2010, p. 626-652. (Landolt-Boernstein
  New Series I, v. 23). (ISBN: 978-3-642-01538-0, 978-3-642-01539-7 (eBook))\/}
  pp 626--652 (\textit{Preprint} \eprint{0810.2520})

\bibitem{cumulant-b}
Athanasiou C, Rajagopal K and Stephanov M 2010 {\em Phys. Rev. D\/} {\bf 82}
  074008 (\textit{Preprint} \eprint{1006.4636})

\bibitem{correlation-koch}
Bzdak A, Koch V and Strodthoff N 2017 {\em Phys. Rev. C\/} {\bf 95} 054906
  (\textit{Preprint} \eprint{1607.07375})

\bibitem{correlation-Lin}
Lin Y, Chen L and Li Z 2017 {\em Phys. Rev. C\/} {\bf 96} 044906
  (\textit{Preprint} \eprint{1707.04375})

\bibitem{star-prl-1}
Adamczyk L {\em et~al.\/} (STAR) 2014 {\em Phys. Rev. Lett.\/} {\bf 112} 032302
  (\textit{Preprint} \eprint{1309.5681})

\bibitem{star-prl-2}
Adamczyk L {\em et~al.\/} (STAR) 2014 {\em Phys. Rev. Lett.\/} {\bf 113} 092301
  (\textit{Preprint} \eprint{1402.1558})

\bibitem{Xue-JPG42}
Pan X, Xu M and Wu Y 2015 {\em J. Phys. G\/} {\bf 42} 015104 (\textit{Preprint}
  \eprint{1407.7957})

\bibitem{Zhonghs-Luoxf-2016}
Fan W, Luo X and Zong H 2017  (\textit{Preprint} \eprint{1702.08674})

\bibitem{finite-size-a}
Weber C, Capriotti L, Misguich G, Becca F, Elhajal M and Mila F 2003 {\em Phys.
  Rev. Lett.\/} {\bf 91}(17) 177202

\bibitem{finite-size-b}
Olsson P 1997 {\em Phys. Rev. B\/} {\bf 55}(6) 3585--3602

\bibitem{brazil}
Palhares L~F, Fraga E~S and Kodama T 2011 {\em J. Phys. G\/} {\bf 38} 085101
  (\textit{Preprint} \eprint{0904.4830})

\bibitem{PRD90-2014}
Tripolt R~A, Braun J, Klein B and Schaefer B~J 2014 {\em Phys. Rev. D\/} {\bf
  90} 054012 (\textit{Preprint} \eprint{1308.0164})

\bibitem{PLB713-2012}
Braun J, Klein B and Schaefer B~J 2012 {\em Phys. Lett. B\/} {\bf 713} 216--223
  (\textit{Preprint} \eprint{1110.0849})

\bibitem{Wu-Lizhu}
Wu Y, Chen L and Chen X~S 2009 {\em PoS\/} {\bf CPOD2009} 036

\bibitem{Lizhu-wu}
Lizhu C, Chen X~S and Yuanfang W 2009  (\textit{Preprint} \eprint{0904.1040})

\bibitem{shift-b}
Privman V (ed) 1990 {\em {Finite size scaling and numerical simulation of
  statistical systems}\/}

\bibitem{scaling}
Cardy J~L 1996 {\em {Scaling and renormalization in statistical physics}\/}

\bibitem{RoyPRL}
Lacey R~A 2015 {\em Phys. Rev. Lett.\/} {\bf 114} 142301 (\textit{Preprint}
  \eprint{1411.7931})

\bibitem{Roy1}
Lacey R~A 2016 {\em Nucl. Phys. A\/} {\bf 956} 348--351 (\textit{Preprint}
  \eprint{1512.09152})

\bibitem{Roy2}
Lacey R~A, Liu P, Magdy N, Schweid B and Ajitanand N~N 2016  (\textit{Preprint}
  \eprint{1606.08071})

\bibitem{PRD97-NA49}
Antoniou N~G, Diakonos F~K, Maintas X~N and Tsagkarakis C~E 2018 {\em Phys.
  Rev. D\/} {\bf 97} 034015 (\textit{Preprint} \eprint{1705.09124})

\bibitem{Lizhu-Wucpod11}
Chen L, Ming S, XSChen and Wu Y 2011 {\em CPOD\/}

\bibitem{Fisher:1982xt}
Fisher M~E and Berker A~N 1982 {\em Phys. Rev.\/} {\bf B26} 2507--2513

\bibitem{vanLeeuwen:1975zz}
van Leeuwen J~M~J 1975 {\em Phys. Rev. Lett.\/} {\bf 34} 1056--1058

\bibitem{PhysRevLett.35.477}
Nienhuis B and Nauenberg M 1975 {\em Phys. Rev. Lett.\/} {\bf 35}(8) 477--479

\bibitem{2D-ising-a}
McCoy B and Wu T 1973 {\em The two-dimensional Ising model\/} (Harvard
  University Press)

\bibitem{2D-ising-b}
Landau D~P 1976 {\em Phys. Rev. B\/} {\bf 13}(7) 2997--3011

\bibitem{2D-ising-c}
Newman M and Barkema G 1999 {\em Monte Carlo Methods in Statistical Physics\/}
  (Oxford University Press: New York, USA)

\bibitem{potts-model-wu}
Wu F~Y 1982 {\em Rev. Mod. Phys.\/} {\bf 54} 235--268 [Erratum: Rev. Mod.
  Phys.55,315(1983)]

\bibitem{potts-Z-a}
DeGrand T~A and DeTar C~E 1983 {\em Nucl. Phys. B\/} {\bf 225} 590

\bibitem{potts-Z-b}
Polyakov A~M 1978 {\em Phys. Lett. B\/} {\bf 72} 477--480

\bibitem{potts-Z-c}
Susskind L 1979 {\em Phys. Rev. D\/} {\bf 20} 2610--2618

\bibitem{potts-Z-d}
Patel A 1984 {\em Nucl. Phys. B\/} {\bf 243} 411--422

\bibitem{potts-first-order-a}
Bl\"ote H~W~J and Swendsen R~H 1979 {\em Phys. Rev. Lett.\/} {\bf 43}(11)
  799--802 \urlprefix\url{https://link.aps.org/doi/10.1103/PhysRevLett.43.799}

\bibitem{potts-first-order-b}
Janke W and Villanova R 1997 {\em Nucl. Phys. B\/} {\bf 489} 679--696
  (\textit{Preprint} \eprint{hep-lat/9612008})

\bibitem{potts-ising-b}
Karsch F and Stickan S 2000 {\em Phys. Lett. B\/} {\bf 488} 319--325
  (\textit{Preprint} \eprint{hep-lat/0007019})

\bibitem{HBT-1}
Rischke D~H and Gyulassy M 1996 {\em Nucl. Phys. A\/} {\bf 608} 479--512
  (\textit{Preprint} \eprint{nucl-th/9606039})

\bibitem{HBT-2}
Csorgo T and Lorstad B 1996 {\em Phys. Rev. C\/} {\bf 54} 1390--1403
  (\textit{Preprint} \eprint{hep-ph/9509213})

\bibitem{HBT-3}
Chapman S, Scotto P and Heinz U~W 1995 {\em Phys. Rev. Lett.\/} {\bf 74}
  4400--4403 (\textit{Preprint} \eprint{hep-ph/9408207})

\bibitem{Book-phase}
Domb C, Green M~S and Lebowitz J~L 1991 {\em Phase Transitions and Critical
  Phenomena, Volume 14\/} (Academic Press)

\bibitem{Engels:2011km}
Engels J and Karsch F 2012 {\em Phys. Rev.\/} {\bf D85} 094506
  (\textit{Preprint} \eprint{1105.0584})

\bibitem{Blote_1995}
Blöte H~W~J, Luijten E and Heringa J~R 1995 {\em Journal of Physics A:
  Mathematical and General\/} {\bf 28} 6289--6313 (\textit{Preprint}
  \eprint{cond-mat/9509016v1})

\end{thebibliography}

\end{document}